\documentstyle[psfig,epsfig,aps,aps12]{revtex}
\def\be{\begin{equation}}
\def\ee{\end{equation}}
\def\bea{\begin{eqnarray}}
\def\eea{\end{eqnarray}}
\def\lbl{\label}

\begin{document}
\draft
\title{Brane {\it versus} shell cosmologies in Einstein and Einstein-Gauss-Bonnet theories}
\author{Nathalie Deruelle$^{1,2,3}$ and Tom\'a\v s Dole\v zel$^{1,4}$}
 \address {$^1$ D\'epartement d'Astrophysique Relativiste et de Cosmologie,\\ 
UMR 8629 du Centre National de la Recherche Scientifique,\\ 
Observatoire de Paris, 92195 Meudon, France}
 \address{$^2$ Institut des Hautes Etudes Scientifiques,\\ 
91140 Bures-sur-Yvette, France}
 \address{$^3$ Centre for Mathematical Sciences, DAMTP,\\ 
University of Cambridge, Wilberforce Road, Cambridge, CB3 0WA, England}
\address{$^4$ Institute of Theoretical Physics, Charles University,\\ 
V Hole\v sovi\v ck\'ach 2, 18000 Prague 8, Czech Republic}
\date{\today}

\maketitle

\begin{abstract}
We first illustrate on a simple example how, in existing brane cosmological models, the connection 
of a 'bulk' region to its mirror image creates matter on the 'brane'.
Next, we present a cosmological model with no $Z_2$ symmetry which is a spherical symmetric
'shell' separating two {\it metrically  different} 5-dimensional anti-de Sitter regions. We find
that our model becomes Friedmannian at late times, like present brane models, but that its early time behaviour is
very different: the scale factor grows from a non-zero value at the big bang singularity. 
We then show how the Israel matching conditions across the membrane (that is either a brane or a shell)
have to be modified if more general equations than Einstein's, including a Gauss-Bonnet
correction, hold in the bulk, as is likely to be the case in a low energy limit of string theory. We find that the
membrane can then no longer be treated in the thin wall approximation. However its microphysics may, in some
instances, be simply hidden in a renormalization of Einstein's constant, in which cases Einstein and Gauss-Bonnet
membranes are identical.
\end{abstract}

\pacs{ 98.80.Cq, 98.70.Vc}


\section*{I Introduction}

Motivated by recent developments in high energy physics [12,16,17,18] there is at present a considerable
increase of activity in the domain of cosmology with extra dimensions. 
In these models
gravity is assumed to act in a $n$-dimensional 'bulk' while the standard
model interactions are confined to a 4-dimensional slice ('brane' worldsheet) of this multi-dimensional
spacetime. Randall and Sundrum
have recently proposed two models in which all the matter  
is confined to a 4-dimensional brane worldsheet embedded in a 5-dimensional 
anti-de Sitter ($AdS_5$) spacetime with imposed $Z_2$ metric symmetry, $w\to-w$ ($w$ denotes the fifth dimension) [13].
In the first model, the fifth dimension 
is compact and bounded by two branes with positive and negative tension, the visible universe 
being modelled by the negative tension brane.  
Their second model assumes a single brane embedded
in a $AdS_5$ with a non-compact fifth dimension.

More recently, several authors have found exact cosmological solutions describing
'4+1' brane universes. Bin\'etruy et al. [1,2] solved the Einstein equations for a single brane embedded in
a bulk governed by a negative cosmological constant. 
They impose the $Z_2$ symmetry to obtain a global cosmological solution
in the Gaussian normal coordinates ('BDL' solutions). Later, Ida [3] and Kraus [15] considered in Schwarzschild  
coordinates a spherically symmetric hypersurface moving in a
5-dimensional Schwarzschild-$AdS_5$ spacetime. By performing an explicit coordinate transformation from 
the Schwarzschild coordinates used in [3,15]
to the Gaussian normal coordinates used in [1,2], Mukhoyama et al. [4] showed that all these solutions represent the same spacetime
described in different coordinate systems. 

Now, what is usually done in General Relativity, when, e.g., studying the gravitational
collapse
of spherical bodies is to join two {\it metrically different} solutions of the Einstein field equations.  
Integration of the Einstein 
equations across the surface separating the two regions leads to the Israel junction conditions ([5], see
also Appendix A for a review) relating the surface stress-energy tensor to the discontinuity of the extrinsic curvature
across the surface. The sign of the extrinsic curvature, and thus the form of the junction
conditions, depend on the definitions of normal vectors in the neighbourhood of the surface. However, once the
directions of the normal vectors on each side of the surface are fixed, e.g. pointing in a defined positive sense, 
the formalism becomes unambiguous and yields in general
a nonzero stress-energy tensor for the surface (massive 'shell'). 

On the other hand matter of $Z_2$ symmetric $AdS_5$ branes, which connect two {\it metrically identical} solutions of the Einstein equations, arises
formally by a flip of the normal vectors at one or the other side of the boundary surface while
preserving the form of the Israel junction conditions.  
We illustrate in section II how this flip of the normals across the brane 
describes the connection, via the Israel junction conditions, of a bulk region 
and its mirror image. Were no formal flip of the normal vectors performed, i.e. dropping out of the $Z_2$ symmetry,
one would join in a topologically trivial way 
two complementary parts of the $AdS_5$ spacetime across a non-massive boundary surface. 

As pointed out by Bin\'etruy et al., brane cosmology leads to Friedmann-like equations different from the standard ones.
They showed, however, that the standard cosmological evolution can be obtained in a $Z_2$ symmetric 
model \`a la Randall-Sundrum if one finely
tunes the tension of the brane. In section III we show that the standard behaviour at late times can also beobtained when
one drops out the $Z_2$ symmetry: we present a cosmological model which is a shell 
separating two metrically different anti-de Sitter regions.
Just as Bin\'etruy et al. we fine-tune the tension of the shell
(that is we impose a particular equation of motion for the shell) 
in order to recover standard cosmology at late times in the shell.
On the other hand the evolution of the scale factor at early times is very different from what give
standard as well as brane cosmologies, as the scale factor grows from a non-zero value at the big bang singularity.

If these brane or shell models are to be the low energy limit of string theory, it is likely that the field
equations include in particular the Gauss-Bonnet term, which, in five dimensions, is the only
non-linear term in the curvature which yields second order field equations (see e.g. [10] and references therein).
In section IV we show how the Israel matching conditions across the membrane
(either brane or shell) have to be modified to take into account 
the Gauss-Bonnet correction.
We conclude that in this case the thin membrane approximation fails. The complete evolution of the Einstein-Gauss-Bonnet 
universe must be studied more carefully taking into account the internal structure of the membrane.
We shall see however that the microphysics of the thick membrane may, at late times, be hidden in a renormalisation of Newton's 
constant so that the Einstein and Einstein-Gauss-Bonnet membranes become identical.

\section*{II The geometry and topology of the BDL brane solutions}

In this section we analyse the first BDL brane model [1], which
shares the same topological properties as the models discussed in [2,3,4,15] and, although simpler, 
is not included
in the analysis of Ida [3] and Mukhoyama et al.[4].
Consider, in a coordinate system $x^A=\{\tau,x^i,w\}$ ($i=1,2,3$), 
the 5-dimensional metric of the first BDL brane model
\be ds^2=-n^2(\tau,w)d\tau^2+S^2(\tau,w)\delta_{ij}dx^idx^j+dw^2\lbl{2.1}\ee
with
\bea S(\tau,w)&=&a(\tau)-\dot a(\tau)|w|\nonumber\\
     n(\tau,w)&=&1-{\ddot a(\tau)\over \dot a(\tau)}|w|\nonumber\eea
where $a(\tau)$ is an arbitrary function of time $\tau$, where a dot denotes a time derivative and
$|...|$ an absolute value.

It is easy to check that the Riemann tensor of the metric (\ref{2.1}) is zero everywhere, except on the brane $\Sigma$
defined by $w=0$, where it exhibits a $\delta$-like discontinuity. Now, were the $Z_2$ symmetry of the metric dropped
out and $|w|$ replaced by $w$ or $-w$,  the metric
(\ref{2.1}) would be  flat everywhere, including on $\Sigma$. The presence of matter on the brane is therefore due to the
reflexion symmetry imposed on the metric.

In order to illustrate how matter on the brane arises from a topologically non trivial pasting of two flat regions, we
transform the metric  (\ref{2.1}) into its minkowskian form
\be ds^2=-(dX^0)^2+(dX^1)^2+(dX^2)^2+(dX^3)^2+(dX^4)^2.\lbl{2.2}\ee
A simple way to obtain the transformation $X^B=X^B(x^A)$ is to integrate the equations which give the Christoffel
symbols $\Gamma^A_{BC}$ for the metric (\ref{2.1}) as
\be{\partial X^A\over\partial x^E}\Gamma^E_{BC}
={\partial^2X^A\over\partial x^B\partial x^C}\lbl{2.3}\ee
and to impose
\be g_{AB}={\partial X^C\over\partial x^A}
{\partial X^D\over\partial x^B}\eta_{CD}.\lbl{2.4}\ee
After some algebra, one gets (up to a Lorentz transformation)~:
\bea X^0&=&S(\tau,w)\left({r^2\over4}+1-{1\over4h^2a^2}\right)-{1\over2}\int\! d\tau {\dot h\over
ah^3}\nonumber\\
     X^i&=&S(\tau,w)\,x^i\lbl{2.5}\\
     X^4&=&S(\tau,w)\left({r^2\over4}-1-{1\over4h^2a^2}\right)-{1\over2}\int\! d\tau{\dot h\over ah^3}.\nonumber\eea
where $h\equiv \dot a/a$ and $r^2\equiv\delta_{ij}x^ix^j$.

Inverting (\ref{2.5}) gives $|w|$ as some function of the $X^A$  coordinates~: $|w|=f(X^A)$. In the simple case
$a(\tau)=\tau$, we have for example
\be |w|={1\over4}(3X^0+5X^4)-{R^2\over X^0-X^4}\lbl{2.6}\ee
(with $R^2\equiv\delta_{ij}X^iX^j$). As for the brane $\Sigma$, it is represented by the hypersurface
$f(X^A)=0$. If $a(\tau)=\tau^q$ for example, its equation reads
\be{\eta_{AB}X^AX^B\over(X^0-X^4)^{2q}}=-{q^2\over2^{2q}(2q-1)}.\lbl{2.7}\ee

The point to note is that the coordinates $x^A =\{\tau,x^i,w\}$ cover only the portion $f(X^A)\geq 0$ of the full Minkowski
spacetime spanned by the $X^A$  coordinates, and that each point of this 'half'-space represents two points of the bulk, one
corresponding to, say, $w=a$ and the other to $w=-a$. In order to have a one-to-one correspondence between the $x^A$
and the $X^A$ coordinates, one  must unfold this half Minkowski space along the brane $\Sigma$. In doing so the normal
vector $n^A$, chosen to point to, say, increasing $f(X^A)$, will point to positive, increasing, $w$ on one side of the brane
(the '$+$' side), and to negative, decreasing, $w$ on the '$-$' side. In other words, $n^A$ flips across the brane.  As a
consequence the extrinsic curvature $K^\pm_{\mu\nu}$ of the $w\to0_\pm$ surfaces ($x^\mu$ being four coordinates on
the brane and $ds^2_\Sigma=\gamma_{\mu\nu}dx^\mu dx^\nu$ its induced metric) becomes discontinuous across the
brane in such a way that
\be K_{\mu\nu}^+=-K_{\mu\nu}^-.\lbl{2.8}\ee
The standard Israel junctions conditions then give the stress-energy tensor ${\cal
T}_{\mu\nu}$ of the matter on the brane as
\be \kappa \left({\cal T}_{~\nu}^\mu-{1\over3}\delta_{~\nu}^\mu{\cal
T}\right)=\hat K_{~\nu}^\mu\quad\hbox{with}\quad \hat K_{\mu\nu}\equiv  K_{\mu\nu}^+-K_{\mu\nu}^-,\lbl{2.9}\ee
where indices are raised with the inverse metric $\gamma^{\mu\nu}$,  where
${\cal T}\equiv\gamma^{\mu\nu}{\cal T}_{\mu\nu}$ and where $\kappa$ is Einstein's constant.

If one had simply considered, in a flat, simply oriented, 5-dimensional universe in Minkowskian coordinates $X^A$, the
surface
$\Sigma$ defined by e.g. (\ref{2.6}), there would have been
(with our convention on the directions of normal vectors) 
no discontinuity in the extrinsic curvature across that surface
so that no matter would have been required on
$\Sigma$ for the Israel matching conditions (\ref{2.9}) to be satisfied. Equivalently, as we have already
remarked, if one replaced in the BDL metric (\ref{2.1}) $|w|$ by $w$ or $-w$, 
the coordinates $x^A$ would cover the full Minkowski spacetime
and one would describe a completely flat universe, 
so that there would be no discontinuity in the extrinsic curvature across the surface $w=0$ and no matter on the brane. 

The procedure to obtain (\ref{2.9}) is reminiscent of what is sometimes done in the Kerr spacetime, where a region between
two planes $z=\pm z_0$ is excised in order to create a discontinuity in the extrinsic curvature and hence a source for the 
Kerr solution (cf. e.g. [9]).

The more general, second, BDL brane cosmological model [2] is a solution, everywhere except on the brane $\Sigma$,  of
$G_{AB}+\Lambda g_{AB}=0$, where
$G_{AB}$ is the five-dimensional Einstein tensor of the metric $g_{AB}$ and $\Lambda (<0)$ a cosmological constant.
Mukhoyama [4] et al. showed that the bulk metric is that of a
Schwarzschild-anti-de Sitter space time: they performed a coordinate transformation
$\{\tau, x^i,w\}\to\{t,r,\chi,\theta,\phi\}$ in order to bring the BDL metric in the bulk into the more familiar form
\be ds^2=-\Phi(r)\, dt^2+{dr^2\over\Phi(r)}+r^2[d\chi^2+f^2_k(\chi)(d\theta^2+\sin^2\theta d\phi^2)]\lbl{2.10}\ee
with
\be\Phi(r)=k-{M\over r^2}-{\Lambda\over6} r^2\ee
where $f_1(\chi)=\sin\chi$, $f_0(\chi)=\chi$, $f_{-1}(\chi)=\sinh\chi$, and where $M$ is the Schwarzschild mass
parameter. (Note that the case $k=M=\Lambda=0$ considered above is not included in this transformation.) In the
coordinates $\{t,r,\chi,\theta,\phi\}$ the  brane $\Sigma$ is a four dimensional sphere, $r=a(t)$ (and hence 
is geometrically simpler than when the bulk is flat, cf. e.g. (\ref{2.7})). 

Again, the presence of matter on $\Sigma$ arises from the fact that the BDL spacetime is not a simply oriented
Schwarzschild-anti-de Sitter space time but is obtained by disregarding the region, say, inside the hypersphere $\Sigma$
and unfolding the region outside along it, which implies a flip of the normal vectors across $\Sigma$.
In a simply oriented spacetime (\ref{2.10}) there would be no need for the matter 
on the hypershpere $\Sigma$ for the junction conditions to be satisfied. 

\section*{III The universe as a shell separating two anti-de Sitter regions}

Let us now consider, in a spirit more akin to what is usually done in General Relativity, a simply
oriented five dimensional spacetime, consisting of two metrically different anti-de Sitter regions, one solution of
$G_{AB}+\Lambda_+ g_{AB}=0$, the other of $G_{AB}+\Lambda_- g_{AB}=0$, separated by a shell
$\Sigma$.
The bulk regions, with Riemann tensors
$R_{ABCD}=L_{\pm}(g_{AC}g_{BD}-g_{AD}g_{BC})$, where $L_{\pm}=\Lambda_{\pm}/6$, are both described
by coordinates $(t,r,\chi,\theta,\phi)$. 
The metrics outside, resp. inside, $\Sigma$ have the following forms  
\bea ds^2|_+&=&-\Phi_+(r)dt^2+\Phi_+^{-1}(r)dr^2+r^2[d\chi^2+f_k^2(\chi)
(d\theta^2+\sin^2\theta d\phi^2)]\,,\lbl{31}\\ 
     ds^2|_-&=&-\Phi_-(r)\alpha^2(t)dt^2+\Phi_-^{-1}(r)dr^2+
r^2[d\chi^2+f_k^2(\chi)(d\theta^2+\sin^2\theta d\phi^2)],\lbl{33}
\eea 
with
\be \Phi_{\pm}(r)=k-L_{\pm}r^2\lbl{34},\ee
and where the lapse function $\alpha(t)$ is introduced in order to match the two coordinate grids through the shell
\footnote{One can consider two Schwarzschild-Anti de Sitter regions 
corresponding to the BDL solution with $\Lambda<0$, ${\cal C} \neq 0$ [2]. Ida [3] showed that 
the Schwarzschild mass parameter $M=M_+=M_-$ generates an effective radiative term in the 
Friedmann-like equation, $\propto Ma^{-4}$, and thus corresponds to the constant ${\cal C}$ of
the BDL solution. We shall not consider this generalisation here (cf. also [19]).}.

The equation of motion of the shell is given parametrically by
\be r=a(\tau)\quad;\quad t=t(\tau).\ee  
Choosing $\tau$ as the proper time on the shell and requiring 
the continuity of the metrics (\ref{31}) and (\ref{33}) at $\Sigma$ then gives 
\be \dot t = {\sqrt{\Phi_+(a)+\dot a^2}\over \Phi_+(a)}\quad;\quad 
\alpha(\tau)={\Phi_+(a)\over\Phi_-(a)}\sqrt{\dot a^2 + \Phi_-(a) \over 
\dot a^2 + \Phi_+(a)},\lbl{37}\ee
where a dot denotes $d/d\tau$.
The metric on $\Sigma$ then takes the Friedmann-Lema\^itre form
\be ds^2_{\Sigma}\equiv \gamma_{\mu\nu} dx^{\mu}dx^{\nu}=-d\tau^2+a^2(\tau)
[d\chi^2+f_k^2(\chi)(d\theta^2+\sin^2\theta d\phi^2)].\lbl{38}\ee

In the interior and exterior regions,
the independent tangent vectors at the shell are given as 
$e^{A ~\pm}_{\tau}=(\dot t,\dot a,0,0,0)$ and
$e^{A ~\pm}_{i}=\delta^A_i$ where the index $i$ stands for $(\chi, \theta, \phi)$.
The normal vectors to the shell both pointing in the positive direction, by definition from '$-$' to
'$+$', normalized to unity, are
$$n_A^-=\alpha(t)(-\dot a,\dot t,0,0,0) \quad,\quad n_A^+=(-\dot a,\dot t,0,0,0).$$
The extrinsic curvature is defined by
$$K_{\mu\nu}^{\pm}\equiv -e_{\mu}^A e_{\nu}^B\nabla_A n_B|^{\pm},$$
where the indices $(\mu,\nu)$ stand for $(\tau,\chi,\theta,\phi)$ and are raised and lowered by the induced 
Friedmann-Lema\^itre metric (\ref{38}). $\nabla_A$ is the covariant derivative associated with the metric (\ref{31})
or (\ref{33}).

The non-vanishing components of the extrinsic curvature read 
\bea K_{\tau}^{\tau ~\pm}&=&-{1\over\sqrt{h^2+{k\over a^2}-L_{\pm}}}
\left[{\ddot a\over a} - L_{\pm} \right],\nonumber\\
\quad K_{\chi}^{\chi ~\pm}&=&K_{\theta}^{\theta ~\pm}=
K_{\phi}^{\phi ~\pm}=-\sqrt{h^2+{k\over a^2}-L_{\pm}}\,,
\lbl{39}\eea
with $h\equiv \dot a/a$.

Israel's junction conditions 
give the stress-energy tensor of the shell in terms of the jump in its extrinsic curvature.
The stress-energy tensor of the shell has the perfect-fluid form, in the coordinates of metric (\ref{38}) it reads
${\cal T }^{\mu}_{\nu}=diag(-\rho,p,p,p)$ where $\rho$ and $p$ are the energy density and pressure of the
shell respectively. The junction condition (\ref{2.9}) gives 
\be \rho = {3\over \kappa} \hat K^{\chi}_{~\chi}.\lbl{41}\ee
As for $p$, it is given by the conservation law 
\be \dot\rho+3(\dot a/a)(\rho+p)=0.\lbl{ZZ}\ee 
Putting together (\ref{39}) and (\ref{41}) we get 
\be {\kappa\rho\over 3}=\sqrt{h^2+{k\over a^2}-L_-}-\sqrt{h^2+{k\over a^2}-L_+}.\lbl{42}\ee

By contrast the brane cosmological model is obtained by imposing the $Z_2$ symmetry: 
$L\equiv L_+=L_-$ and changing the 
direction of one of the normal vectors. The reflexion $n^A_+=-n^A_-$ then gives in the brane
\be {\kappa\rho\over 6}=\sqrt{h^2+{k\over a^2}-L}\,\lbl{42a}.\ee
In 'shell cosmology' on the other hand to impose  $L_+=L_-$ would yield a massless boundary surface.

In order to recover standard cosmology at late times in the shell we
have to consider $L_+\leq 0$, $L_-\leq 0$, otherwise $h^2+k/a^2$ can never go to zero. We also choose 
$L_+<L_-<0$ (the particular case $L_-=0$ is considered below). 
Furthermore, as in [2], we must  decompose the stress-energy tensor in the shell into:
\be \rho\equiv \rho_m +\sigma\quad;\quad p\equiv p_m-\sigma,\lbl{DEC}\ee
where $\rho_m$, $p_m$ can be interpreted as the energy density and pressure of ordinary matter and where the tension
$\sigma$ must be fine-tuned as
$$\sigma\equiv -{3\over\kappa}\left(\sqrt{-L_+}-\sqrt{-L_-}\right).$$
Equation (\ref{42}) then reduces at late times to the Friedman-like equation for the energy density of 
ordinary matter
\be h^2+{k\over a^2}={8\pi G\over 3}\rho_m+O(\rho_m^2),\lbl{t1}\ee
where
\be {8\pi G\over 3}\equiv {2\kappa\over 3}{\sqrt{L_+L_-}\over\sqrt{-L_+}-\sqrt{-L_-}}>0,\lbl{t2}\ee
and $G$ can be interpreted as Newton's constant.
Thus, at late times one recovers as in brane cosmology the standard FL behaviour. 

The complete time evolution of the shell universe can be analysed easily for $k=0$.
The conservation law (\ref{ZZ}) for the equation of state
\be p_m=v\rho_m,\quad v=constant,\ee
gives  $\rho_m\propto a^{-q}$, $q\equiv 3(1+v)$. Proceeding as in standard cosmology, we then
obtain from (\ref{42}) and (\ref{DEC}):
\be \tau={2\over q\sigma}\int^1_{\rho_m\over\sigma}{(1-x)dx \over x^{2/3}
\sqrt{2-x}}{1\over\sqrt{-x^2+2x+\gamma}}
\quad\quad\hbox{with}\quad\quad\gamma\equiv {4\sqrt{L_+L_-}\over\sigma^2}.\lbl{45}\ee
The behaviour of the shell in the neighbourhood of the big bang is different from that 
in both standard $(a\propto \tau^{2/q})$ and brane $(a\propto \tau^{1/q})$ cosmologies. 
Indeed, in the passage to the limit $\tau\to 0$ in (\ref{45}) we have
\be {a\over a_{bb}}\approx 1+{1\over q}\sqrt{\tau\over d}\quad\quad\hbox{with}\quad\quad{1\over d}\equiv
q\sigma\sqrt{1+\gamma}.\ee
Hence, the scale factor grows from a non-zero value $a_{bb}$ at the big bang as $\sim \sqrt{\tau}$ 
(for all equations of state $p_m=v\rho_m$, $v\geq 0$). 
Nevertheless, the singularity of curvature is present at the big bang: 
$R^{\mu}_{\,\,\nu\kappa\lambda}\rightarrow\infty$, because $\dot a,\ddot a\rightarrow\infty$ when $\tau\to 0$. 

In the particular case $L_-=0$ the shell connects a flat and a $AdS_5$ regions. It can therefore represent the 'negative-tension
brane' in the GRS model of 'quasi-localized' gravity [20] (cf. also [21,22]). 
As one can see from (\ref{t1}-\ref{t2}), $G=0$ and standard cosmology
cannot be recovered at late times in such a shell; we have $a\propto\tau^{1/q}$ instead of $a\propto\tau^{2/q}$
when $\tau\to\infty$.

\section*{IV Einstein-Gauss-Bonnet membrane cosmologies}

Up to now we have considered membrane cosmological models which are solutions of the Einstein equations:
$G_{AB}+\Lambda g_{AB}=0$ everywhere except on the membrane $\Sigma$. In brane cosmology the cosmological
constant
$\Lambda $ is the same on each side of $\Sigma$, in shell cosmology, it jumps from $\Lambda_+$ to
$\Lambda_-$. Now, if these models are to be the low energy limit of string theory, it is likely that the field
equations should be generalized and include the Gauss-Bonnet term (cf. e.g. [10] and ref. therein).
In this section we shall therefore consider the gravitational action
\be S_g=\int\! d^5\! x\,\sqrt{-g}(-2\Lambda +R+\alpha L_2)\lbl{4.1}\ee
with
\be L_2=R_{ABCD}R^{ABCD}-4R_{AB}R^{AB}+R^2\ee
where $\alpha$ is a coupling constant, and where $R_{ABCD}$,  $R_{AB}$ and $R$ are the Riemann tensor, the Ricci tensor
and the scalar curvature of the five dimensional metric $g_{AB}$ with determinant $g$.  The corresponding field
equations, outside the membrane, are (see e.g. [10])
\be\Lambda g_{AB}+G_{AB}+\alpha H_{AB}=0\lbl{4.2}\ee
with
\be H_{AB}\equiv 2R_{ALMN}R_B^{\ \ LMN}-4R_{AMBN}R^{MN}-4R_{AM}R_B^{\ \ M}+2RR_{AB}-{1\over2}g_{AB}L_2.\ee

Contrarily to Einstein's equations, the equations (\ref{4.2})  possess, for $\alpha\neq0$ and a given value of the cosmological
constant $\Lambda $, two (anti) de Sitter solutions 
\bea R_{ABCD}&=&L_\pm(g_{AC}g_{BD}-g_{AD}g_{BC})\nonumber\\
\hbox{with}\qquad L_\pm&=&{1\over4\alpha}\left(-1\pm\sqrt{1+{4\alpha\Lambda\over3}}\right).\lbl{4.3}\eea
In brane cosmology we
shall choose one or the other solution everywhere in the bulk. In shell cosmology, we shall choose the
solution
$L_+$ on one side of the shell and the solution $L_-$ on the other side. (Shell cosmologies are therefore more
satisfactory in Einstein-Gauss-Bonnet theory as one does not have to impose different cosmological constants on each
side of the shell.)

In order to get the stress-energy tensor on the membrane, one proceeds along Israel's line.
Like Binetruy et al., we first choose a Gaussian coordinate system $(w,x^\mu)$ such that the metric reads
\bea ds^2&=&dw^2+\gamma_{\mu\nu}dx^\mu dx^\nu\nonumber\\
&=&dw^2-n^2(\tau,w)d\tau^2+S^2(\tau,w)[d\chi^2+f^2_k(\chi)(d\theta^2+\sin^2\theta
d\phi^2)],\lbl{4.4}\eea 
$w=0$ being the equation of the membrane $\Sigma$.
In this coordinate system  the extrinsic curvature of the surfaces $w=constant$ is simply given by
\be K_{\mu\nu}=-{1\over2}{\partial\gamma_{\mu\nu}\over\partial w}.\lbl{4.5}\ee
It jumps across the membrane from $K^+_{\mu\nu}$ to $K^-_{\mu\nu}$ (with $K^+_{\mu\nu}=-K^-_{\mu\nu}$ in the
case of branes) and this discontinuity can be described in terms of the Heaviside distribution.

Expressing now the Riemann tensor (\ref{4.3}) in terms of $K_{\mu\nu}$ and the four dimensional Riemann tensor of the
metric $\gamma_{\mu\nu}$  we then obtain from (\ref{4.2}), everywhere outside the membrane (see Appendices A and B for
the '4+1' decompositions of $G^A_{~B}$ and $H^A_{~B}$)
\be\Lambda \delta_{~\mu}^\nu+G_{~\mu}^\nu+\alpha H_{~\mu}^\nu= (1+4\alpha L)
\left({\partial K_{~\mu}^\nu\over\partial w}-\delta_{~\mu}^\nu{\partial
K\over\partial w}\right)+...\qquad(=0)\lbl{4.6}\ee
 where $K\equiv
\gamma^{\alpha\beta}K_{\alpha\beta}$, where $L=L_+$ or $L_-$, and  where the dots stand for terms containing at most
first order
$w$-derivatives of
$\gamma_{\mu\nu}$.

In Einstein's theory, $\alpha=0$, and (\ref{4.6}), in the vicinity of $\Sigma$, is well defined in a distributional sense:
$\partial K_{~\mu}^\nu/\partial w$ can be expressed in terms of the Dirac distribution and the integration of (\ref{4.6}) across
the membrane gives Israel's junction conditions, that is the stress-energy tensor on the membrane in terms of the
jump in the extrinsic curvature, eq. (\ref{2.9}). 

When $\alpha\neq0$ on the other hand, (\ref{4.6}) is not well defined in a distributional sense, as $L$ cannot be considered as
an infinitely $w$-differentiable function. Indeed,  in shell cosmology, $L$ jumps from $L_+$ to $L_-$ across
$\Sigma$, and in brane cosmology, $L=L_+=L_-$ is continuous across $\Sigma$, but, because of the reflexion
symmetry, has a discontinuous $w$-derivative. This mathematical obstruction simply means that, in
Einstein-Gauss-Bonnet theory, membranes cannot be treated in the thin wall approximation: the jumps in the extrinsic
curvature and in $L$ or its derivative have to be described in detail within  specific microphysical models. 

When the thickness of the  membrane is taken into account, the distributions $\partial K_{~\mu}^\nu/\partial w$ and $L$ 
are replaced by rapidly varying but $C^\infty$ functions. Supposing that the metric keeps the form (\ref{4.4}), we can
define from the
$\tau$-$\tau$ component of (\ref{4.6}) the sharply peaked function (cf. (\ref{BH}) in Appendix B) 
\be\kappa\rho\equiv 3(1+4\alpha L){\partial K_{~\chi}^\chi\over\partial w}\lbl{4.7}\ee
(The $\chi$-$\chi$ component of (\ref{4.6}) is redundant thanks to the conservation
equation, that is the Bach-Lanczos identity (\ref{BI}).)
In the vicinity and inside the membrane $K_{~\chi}^\chi$ can be written as
\be K_{~\chi}^\chi={1\over2}\bar K_{~\chi}^\chi+{1\over2}\hat K_{~\chi}^\chi \,f(\tau,w)\lbl{4.8}\ee
where $\bar K_{\chi\chi}\equiv K^+_{\chi\chi}+K^-_{\chi\chi}$ and  where the function $f(\tau,w)$, which  varies
rapidly from $-1$ to $+1$ across $\Sigma$, encapsulates its microphysics. Similarly, in the case of brane cosmology, we
can write
\be L=\tilde Lg_b(\tau,w)\lbl{4.9}\ee
where $\tilde L=L_+$ or $L_-$ and where $g_b(\tau,w)$ is some even function of
$w$ which varies rapidly from $+1$ to $+1$ across the brane. In shell cosmology on the other hand
\be L={1\over2}\bar L+{1\over2}
\hat L\,g_s(\tau,w)\lbl{4.10}\ee
where $g_s(\tau,w)$ varies rapidly from $-1$ to $+1$ and $\bar L=-{1\over2\alpha}$, $\hat L={1\over2\alpha}\sqrt{1+
{4\alpha\Lambda\over3}}$.

Integrating (\ref{4.7}) across $\Sigma$ we therefore get the energy density of the membrane as
\be\kappa\varrho\equiv\int_{-\eta}^{+\eta}\!dw\,\kappa\rho 
=3\hat K^\chi_{~\chi}\left[1+2\alpha\tilde L\int_{-\eta}^{+\eta}\!dw \,
g_b{\partial f\over\partial w}\right]\lbl{4.11}\ee
in the case of branes, and
\be\kappa\varrho\equiv\int_{-\eta}^{+\eta}\!dw\,\kappa\rho 
={3\over2}\sqrt{1+{4\alpha\Lambda\over3}}\,\hat
K^\chi_{\chi}\int_{-\eta}^{+\eta}\!dw \, g_s{\partial f\over\partial
w}\lbl{4.12}\ee
 in the case of shells (the fact that one does not recover the results of Einstein's theory when $\alpha=0$ is not
surprising as $L_{\pm}$ is divergent in that case) . Now, if $f$ or $g_{b/s}$ depend on $\tau$, the integrals in 
(\ref{4.11})-(\ref{4.12}) are some functions of
$\tau$. But, if $f$ and $g_{b/s}$ do not depend on time, which is probably to be expected whence the brane has reached a
stationary state, that is at late times, then the integrals in (\ref{4.11})-(\ref{4.12}) 
are just  numbers. In this case then, the
microphysics of the membrane is simply hidden in a renormalization of the Einstein constant $\kappa$ and Einstein and
Einstein-Gauss-Bonnet branes are indistinguishable. 

We therefore conclude that Einstein and Einstein-Gauss-Bonnet membranes cosmologies  are probably indistinguishable
(and Friedmannian) at late times, but that the early  time behaviour of the scale factor needs to be studied more
carefully, taking into account the microphysics of the membrane, as is done in e.g. [11]. 


\section*{Acknowledgements}

We thank Thibaut Damour and David Langlois for fruitful discussions.


\appendix

\section{Junction conditions for non-null surfaces in General Relativity}
This appendix summarizes the junction conditions in the theory of Einstein 
(Lanczos [7], Darmois [8], Misner and Sharp [6], Israel [5]). 
Suppose we are given a 4-dimensional hypersurface ($\Sigma$) in a 5-dimensional spacetime (metric $g_{AB}$) 
which can be imagined as the element of a family of surfaces. The normal vectors $n^A$ to this 
family of surfaces are not null; $n_A n^A\equiv \epsilon=\pm 1$. They are all oriented in a
positive direction defined in the bulk. 
Let the surface be either spacelike ($\epsilon=-1$) or timelike ($\epsilon=+1$). 
As an aid in deriving junction conditions we introduce Gaussian normal coordinates 
in the neighbourhood of $\Sigma$. The metric $g_{AB}$ has the form
\be ds^2=\epsilon dw^2+\gamma_{\mu\nu}dx^{\mu}dx^{\nu},\lbl{A1}\ee         
and the extrinsic curvature of the surfaces $w=constant$ is 
\be K_{\mu\nu}=-{1\over 2}{\partial\gamma_{\mu\nu}\over\partial w}.\ee
The curvature tensor of the metric $g_{AB}$ can be expressed in terms of the intrinsic curvature 
of 4-dimensional hypersurface (metric $\gamma_{\mu\nu}$) and of its extrinsic curvature;
one gets the so-called Gauss-Codazzi equations. 
In the special case of Gaussian normal coordinates the equations simplify to
\bea
R_{w\mu w\nu}&=&{\partial K_{\mu\nu}\over \partial w}+K_{\rho\nu}K^{\rho}_{\,\,\mu}\,,\lbl{R1}\\
R_{w\mu\nu\rho}&=&\nabla_{\nu}K_{\mu\rho}-\nabla_{\rho}K_{\mu\nu}\,,\\
R_{\lambda\mu\nu\rho}&=&~^4R_{\lambda\mu\nu\rho}+\epsilon\left[K_{\mu\nu}K_{\lambda\rho}-
                                                                K_{\mu\rho}K_{\lambda\nu}\right]\lbl{R3}\,,
\eea 
where $\nabla_{\rho}$ is the covariant derivative with respect to the 4-dimensional metric $\gamma_{\mu\nu}$. 
From (\ref{R1})-(\ref{R3}) we obtain the decomposition
of the Ricci tensor ($R_{AB}=g^{CD}R_{CADB}$) and of the scalar curvature ($R=g^{AB}R_{AB}$) as:
\bea
R_{ww}&=&\gamma^{\mu\nu}{\partial K_{\mu\nu}\over \partial w}+Tr(K^2)\,,\\
R_{w\mu}&=&\nabla_{\mu}K-\nabla_{\nu}K^{\nu}_{\,\,\mu}\,,\\
R_{\mu\nu}&=&~^4R_{\mu\nu}+\epsilon\left[{\partial K_{\mu\nu}\over\partial w}+2K_{\mu}^{\,\,\rho}K_{\rho\nu}
                                                 -K K_{\mu\nu} \right]\,,\\
R&=&~^4R+\epsilon\left[2\gamma^{\mu\nu}{\partial K_{\mu\nu}\over\partial w}+3Tr(K^2)-K^2\right]\lbl{RK}\,,  
\eea
where we defined $K\equiv K^{\mu}_{\,\,\mu}$ and $Tr(K^2)\equiv K^{\mu\nu}K_{\mu\nu}$. 

In terms of the intrinsic and extrinsic curvature of the 4-dimensional hypersurfaces $w=constant$, the 
Einstein tensor ($G_A^{~B}=R_A^{~B}-(1/2)\delta_A^{~B}R\,$) and the field equations have components
\bea
G^w_{~w}&=&-{1\over 2}~^4R+{1\over 2}\epsilon\left[K^2-Tr(K^2)\right]=\kappa T^w_{~w}\,,\lbl{G1}\\
G^w_{~\mu}&=&\epsilon\left[\nabla_{\mu}K-\nabla_{\nu}K^{\nu}_{\,\,\mu}\right]=\kappa T^w_{~\mu}\,,\lbl{G2}\\
G^{\mu}_{~\nu}&=&~^4G^{\mu}_{~\nu}
              +\epsilon\left[{\partial K^{\mu}_{~\nu}\over\partial w}-\delta^{\mu}_{~\nu}
                   {\partial K\over\partial w}\right]\nonumber\\
              &&+\epsilon\left[- 
                   K K^{\mu}_{~\nu}+{1\over 2}\delta^{\mu}_{~\nu}Tr(K^2)+{1\over 2}\delta^{\mu}_{~\nu} K^2\right]
             =\kappa T^{\mu}_{~\nu}\,.\lbl{G3}
\eea 
If the stress-energy tensor $T^A_{~B}$ contains a 'delta-function contribution' at $\Sigma$, the integral
of $T^A_{~B}$ with respect to the proper distance $w$ measured perpendicularly through $\Sigma$,
\be {\cal T}^A_{~B}\equiv\lim_{\eta\to 0}\left[\int_{-\eta}^{\eta}T^A_{~B}dw\right],\lbl{A3}\ee
is non-zero and represents the surface stress-energy tensor. In this case the extrinsic curvature must be
a distribution of 'Heaviside type' at $\Sigma$. Integral (\ref{A3}) applied on equations (\ref{G1})-(\ref{G3}) yields the 
junction conditions relating the stress-energy tensor of $\Sigma$ to the discontinuity of the extrinsic curvature
at $\Sigma$. In the passage to the limit $\eta\to 0$ only the terms $\sim (\partial K^{\mu}_{~\nu}/\partial w)$ contribute
to yield 
\bea
\kappa {\cal T}^w_{~w}&=&0\,,\nonumber\\
\kappa {\cal T}^w_{~\mu}&=&0\,,\nonumber\\                      
\kappa {\cal T}^{\mu}_{~\nu}&=&\hat K^{\mu}_{~\nu} -\delta^{\mu}_{~\nu}\hat K\,,
\quad\hbox{where}\quad \hat K^{\mu}_{~\nu}\equiv K^{\mu}_{~\nu}(0+)-K^{\mu}_{~\nu}(0-)\lbl{A4}.
\eea
It is useful to denote $K^+_{\mu\nu}\equiv K_{\mu\nu}(0+)$, $K^-_{\mu\nu}\equiv K_{\mu\nu}(0-)$.

As for the intrinsic geometry of $\Sigma$, it must be continuous across $\Sigma$; this is the second junction contition
completing equations (\ref{A4}). If there are no 'delta singularities' contained in $T^A_{~B}$, the bulk is sliced by
massless 'boundary surfaces' (see [5] for details).

\section{Junction conditions for the theory of Einstein-Gauss-Bonnet}
The theory of Einstein-Gauss-Bonnet is based on the following Lagrangian\footnote{We analyse the particular case of 
a 5-dimensional spacetime sliced by 4-dimensional hypersurfaces.}
(cf. e.g. [10] and ref. therein)
\be {\cal L}=\sqrt{-g}\left[-2\Lambda+R+\alpha L_2\right],\ee 
where $g$ is the determinant of the 5-dimensional metric $g_{AB}$ and $\alpha$ is a constant of the dimension 
of $[length]^2$.
$L_2$ is the Gauss-Bonnet Lagrangian which reads
\be L_2=R_{ABCD}R^{ABCD}-4R_{AB}R^{AB}+R^2.\ee
The Euler variation of ${\cal L}$ gives the following field equations:
\be \Lambda g_{AB}+G_{AB}+\alpha H_{AB}=0.\ee
$G_{AB}$ is the Einstein tensor and $H_{AB}$ is its analogue stemmed from the Gauss-Bonnet part of the
Lagrangian, $L_2$,
\bea 
G_{AB}&\equiv& R_{AB}-{1\over2}g_{AB}R\,,\\
H_{AB}&\equiv& 2\left[R_{ALMN}R_B^{\,\,\,LMN}-2R_{AMBN}R^{MN}-2R_{AM}R_B^{\,\,\,M}+RR_{AB}\right]-{1\over 2}
       g_{AB}L_2.\lbl{GB}\quad
\eea
$G_{~B}^A$ and $H_{~B}^A$ satisfy the Bianchi and Bach-Lanczos identities respectively
\be \nabla_A G_{~B}^A=0\quad,\quad\nabla_A H_{~B}^A=0\lbl{BI}.\ee

In order to derive the junction condition in the theory of Einstein-Gauss-Bonnet 
we need to express $H_{AB}$ in terms of the intrinsic curvature of hypersurfaces $w=constant$ and their
extrinsic curvatures. We adopt the notation used in Appendix A in which     
the '4+1' decomposition of the Einstein tensor $G_{AB}$ is shown (expressions (\ref{G1})-(\ref{G3})).

Inserting the decomposition of the curvature tensors (\ref{R1})-(\ref{RK}) into (\ref{GB}) one finds the following 
results: $H^A_{~B}$ does not contain terms $\sim (\partial K^{\mu}_{~\nu}/\partial w)^2$ as one would expect since
$H^A_{~B}$ contains terms $\sim (R_{ABCD})^2$. Further, there are no terms linear in 
($\partial K^{\mu}_{~\nu}/\partial w$) in $H^w_{~w}$ and  $H^w_{~\mu}$ so that these junction 
conditions correspond to those in the theory of Einstein: ${\cal T}^w_{~w}={\cal T}^w_{~\mu}=0$.
The components $H^{\mu}_{~\nu}$ are
\bea
H^{\mu}_{~\nu}&=&\left\{{\partial K^{\mu}_{~\nu}\over \partial w}\right\}\left(2Tr(K^2)-2K^2\right)
                  +\left\{{\partial K^{\mu}_{~\lambda}\over \partial w}\right\}\left(4KK^{\lambda}_{\,\,\nu}
                                                              -4K_{\nu\beta}K^{\beta\lambda}\right)\nonumber\\
                 &+&\left\{{\partial K_{\nu\lambda}\over \partial w}\right\}\left(4KK^{\lambda\mu}
                                                              -4K^{\mu}_{~\beta}K^{\beta\lambda}\right)
                  +\left\{{\partial K \over \partial w}\right\}\left(4K^{\mu}_{~\beta}K^{\beta}_{\,\,\nu}
                                                              -4KK^{\mu}_{~\nu}\right)\nonumber\\
                 &+&\left\{{\partial K_{\alpha\beta}\over \partial w}\right\}\left(4K^{\mu}_{~\nu}K^{\alpha\beta}
                                                              -4K^{\alpha\mu}K^{\beta}_{\,\,\nu}\right)
                  +\left\{{\partial K_{\alpha\beta}\over \partial w}\right\}\delta^{\mu}_{~\nu}
                                                                                   \left(4K^{\alpha}_{\,\,\gamma}
                                                       K^{\gamma\beta}-4KK^{\alpha\beta}\right)\nonumber\\
                 &+&\left\{{\partial K \over \partial w}\right\}\delta^{\mu}_{~\nu}\left(2K^2-2Tr(K^2)\right)\nonumber\\
                 &+&\epsilon\left(-4~^4R_{~\alpha\nu}^{\mu~~~\beta}{\partial K^{\alpha}_{~\beta}\over\partial w}
                                  -4~^4R^{\alpha}_{~\nu}{\partial K^{\mu}_{~\alpha}\over\partial w}
                                  -4~^4R^{\alpha\mu}{\partial K_{\nu\alpha}\over\partial w}\right)\nonumber\\ 
                 &+&\epsilon\left(4~^4R^{\mu}_{~\nu}{\partial K \over\partial w}
                   +2~^4R{\partial K^{\mu}_{~\nu}\over\partial w}
                   +4\delta^{\mu}_{~\nu}~^4R^{\alpha\beta}{\partial K_{\alpha\beta}\over\partial w}                                                 -2\delta^{\mu}_{~\nu}~^4R{\partial K \over\partial w}  \right)\nonumber\\
                 &+&\ldots\,, 
\eea   
where '\ldots' includes terms of zeroth order in $(\partial K^{\mu}_{~\nu}/\partial w)$ which disappear in the passage
to the limit $\eta\to 0$ of the integration (\ref{A3}). 

In what follows we confine the analysis to timelike surfaces ($\epsilon=+1$). The previous expression can be simplified
by using the Leibnitz rule that holds for both the function and the distributions
\bea
H^{\mu}_{~\nu}&=&4{\partial\over\partial w}\left\{KK^{\mu}_{~\alpha}K^{\alpha}_{~\nu}-K^{\mu}_{~\alpha}K^{\alpha\beta}
K_{\beta\nu}+{1\over 2}K^{\mu}_{~\nu}Tr(K^2)- {1\over 2}K^{\mu}_{~\nu}K^2\right\}\nonumber\\
                 &+&4{\partial\over\partial w}\left\{-\delta^{\mu}_{~\nu}{1\over 2}K Tr(K^2)+\delta^{\mu}_{~\nu}
                                             {1\over 3} Tr(K^3)
                 +\delta^{\mu}_{~\nu}{1\over 6}K^3\right\}\nonumber\\
                 &+&4\left(-~^4R_{~\alpha\nu}^{\mu~~~\beta}{\partial K^{\alpha}_{~\beta}\over\partial w}
                                  -~^4R^{\alpha}_{~\nu}{\partial K^{\mu}_{~\alpha}\over\partial w}
                                  -~^4R^{\alpha\mu}{\partial K_{\nu\alpha}\over\partial w}\right)\nonumber\\ 
                 &+&4\left(~^4R^{\mu}_{~\nu}{\partial K \over\partial w}
                   +{1\over 2}~^4R{\partial K^{\mu}_{~\nu}\over\partial w}
                   +\delta^{\mu}_{~\nu}~^4R^{\alpha\beta}{\partial K_{\alpha\beta}\over\partial w}                                                 -{1\over 2}\delta^{\mu}_{~\nu}~^4R{\partial K \over\partial w}  \right)\nonumber\\
                 &+&\ldots\,, 
\eea  
where $Tr(K^3)\equiv K^{\alpha}_{~\beta}K^{\beta}_{~\gamma}K^{\gamma}_{~\alpha}$.
In the case the metric has the form
$$ds^2=dw^2-n^2(\tau,w)d\tau^2+S^2(\tau,w)[d\chi^2+f^2_k(\chi)(d\theta^2+\sin^2\theta d\phi^2)]$$ 
we have
\be H^\tau_{~\tau}=-12L(\tau,w){\partial K^\chi_{~\chi}\over\partial w}+...\quad\hbox{with}\quad
L(\tau,w)\equiv -(K^\chi_{~\chi})^2+{\dot S^2+k n^2\over n^2S^2}\lbl{BH}\ee
and where $K^\chi_{~\chi}=-S'/S$, a prime denotes $\partial/\partial w$. Outside the membrane, spacetime is anti-de Sitter, $n$ and $S$ are given
by BDL [2], and $L(\tau,w)\to L_\pm$ (as an explicit calculation shows). In the vicinity and inside the 
membrane the function $L(\tau,w)$ is either continous with discontinuous
$w$-derivative (case of branes) or discontinous (case of shells) and can be modelled by the expressions 
(\ref{4.9})-(\ref{4.10}) in the text.


\end{document}